\documentclass[lettersize,journal]{IEEEtran}
\usepackage{pifont}
\usepackage{array}
\usepackage{booktabs}
\usepackage{multirow}
\usepackage{subfigure}
\usepackage{times,amsmath,color,amssymb,amsfonts,graphicx,epsfig,cite,psfrag,subfigure,algorithm,epstopdf,bm}
\usepackage{amsmath}
\usepackage{algorithm}
\usepackage{algpseudocode}
\usepackage{textcomp}
\def\BibTeX{{\rm B\kern-.05em{\sc i\kern-.025em b}\kern-.08em
    T\kern-.1667em\lower.7ex\hbox{E}\kern-.125emX}}

\newcommand\eps\varepsilon

\newtheorem{theorem}{\bf{Theorem}}[section]

\begin{document}

%\begin{spacing}{1.5}

\def\FIGDIR{./figs}

\IEEEoverridecommandlockouts

\title{Semantic Extraction Model Selection for IoT Devices in Edge-assisted Semantic Communications}

\author{Hong Chen, \IEEEmembership{Member, IEEE}, Fang Fang, \IEEEmembership{Senior Member, IEEE}, and Xianbin Wang, \IEEEmembership{Fellow, IEEE}
\thanks{Hong Chen and Xianbin Wang are with the Department of Electrical and Computer Engineering, Western University, London, ON N6A 3K7, Canada (e-mail: hche88@uwo.ca; xianbin.wang@uwo.ca).

Fang Fang is with the Department of Electrical and Computer Engineering and also with the Department of Computer Science, Western University, London, ON N6A 3K7, Canada (e-mail: fang.fang@uwo.ca).

%Copyright (c) 2023 IEEE. Personal use of this material is permitted. However, permission to use this material for any other purposes must be obtained from the IEEE by sending a request to pubs-permissions@ieee.org.
}
}

\maketitle \thispagestyle{empty}

\begin{abstract}

Semantic communications offer the potential to alleviate communication loads by exchanging meaningful information. However, semantic extraction (SE) is computationally intensive, posing challenges for resource-constrained Internet of Things (IoT) devices. To address this, leveraging computing resources at the edge servers (ESs) is essential. ESs can support multiple SE models for various tasks, making it crucial to select appropriate SE models based on diverse requirements of IoT devices and limited ES computing resources.
In this letter, we study an SE model selection problem, where the ES co-located at the access point can provide multiple SE models to execute the uploaded SE tasks of associated IoT devices. We aim to maximize the total semantic rate of all SE tasks by selecting appropriate SE models, while considering SE delay and ES capacity constraints, and SE accuracy requirements. The formulated NP-complete integer programming problem is transformed into a modified Knapsack problem. The proposed efficient approximation algorithm using dynamic programming can yield a guaranteed near-optimum solution. Simulation results demonstrate the superior performance of proposed solution.
\end{abstract}

\begin{IEEEkeywords}
\noindent Semantic communications, edge computing, semantic extraction model selection, diverse requirements.
\end{IEEEkeywords}

\sloppy
\allowdisplaybreaks

\vspace{-0.3cm}
\section{Introduction}
\label{sec:introduction}

%(Outlines:
%
%1. What is semantic communications? What advantages it has? what is SE? what weakness it has? IoT devices usually equipped with universal model. not enough to support diverse requirements. However, IoT devices too small to support multiple models.
%
%2. Thus, edge computing can help. Due to the short distance between devices and AP. Raw data can be transmitted to the ES. ES helps to do the SE and forwards to the receiver. ESs train and support multiple SE models for different classes that can result in differing levels of SE accuracy and semantic transmission rate. However, computation capability is limited to serve a large set of IoT devices. How to provide best performance of SE for multiple devices under constrained resources and SE delay and accuracy requirements. Some references: SemCom for MEC, only one model used. MEC for Semcom, only considers computation requirements not accuracy and semantic rate.
%
%3. what we considered in the system. summary of our contribution.
%)

\IEEEPARstart{W}{ith} growing emerging applications and increasing data loads, the bottleneck of spectrum scarcity motivates a paradigm shift from conventional to semantic communications. As a novel paradigm focusing on the meaning of information rather than source data, semantic communications have revealed a significant potential to alleviate the communication loads over the network \cite{yang2023}. As a critical step, semantic extraction (SE) is to extract semantic information from the source data, which is computationally intensive \cite{luo2022semantic,Xie2022Jsac}. For example, the work in \cite{cang2023online} used an universal model at devices, in which only simple background knowledge can be semantically extracted. Due to the limited capacity of internet of things (IoT) devices, it is challenging to implement the SE at device to satisfy diverse quality of service (QoS) requirements of SE tasks.

To realize semantic communications for IoT devices, it is indispensable to exploit all available computing resources at the edge servers (ESs) at access points (APs) \cite{EdgeIntellegence}. Benefit from the proximity from the ES to IoT devices, the generated raw data can be transmitted to the ES in low latency to execute the SE \cite{semanticaware,computing}. In \cite{computing}, the SE task oriented resource allocation in mobile edge computing systems is investigated. However, the authors in \cite{computing} only considered one SE model at the ES without considering diverse semantic requirements. A large number of work \cite{textsemantic,Xie2022WCL} is dedicated to design an effective SE model to achieve high SE accuracy without considering the required computing resources.
%The allocation of multiple SE models for different tasks is not considered yet.
%to meet various requirements
%in edge-assisted semantic-aware networks

Fortunately, ES has the capability to support multiple SE models simultaneously. In general, distinct SE models result in differing levels of SE accuracy and semantic rate requiring different amount of computation load. IoT devices with higher SE accuracy typically use models that require higher levels of computation at the ES \cite{wang2020machine}. However, the competition for constrained ES execution capacity among a large set of IoT devices can degrade the achievable semantic communication performance. How to provide best performance by selecting appropriate SE models under limited edge computing resource while meeting the SE requirements of IoT devices is an urgent issue to be addressed.

Motivated by the above discussion, in this letter, we consider a common scenario that the co-located ES at the AP can support multiple SE models for diverse task classes associated with IoT devices.
%Each model is characterized by its required amount of computational load at the ES, the achievable SE accuracy and the resulted semantic transmission rate.
%Specifically, a semantic extraction model selection problem is studied, where the SE tasks of IoT devices are uploaded to the associated ES to be executed.
The main contributions are as follows:

\begin{itemize}

\item A novel SE model selection problem is studied in an edge-assisted semantic-aware network, where the SE tasks of IoT devices are uploaded to the associated ES for execution. The objective is to maximize the total semantic rate of all SE tasks by selecting appropriate SE models, while considering SE delay and accuracy requirements of IoT devices, and the maximum ES execution capacity.

\item The formulated problem is an NP-complete integer programming, which is transformed into a modified Knapsack problem. The proposed approximation algorithm using dynamic programming (DP) can achieve a solution with guaranteed near-optimum performance in polynomial time.

%which is computationally hard to solve exactly. It is observed that it can be

\item Simulation results are presented to demonstrate the validity and superior performance of proposed solution. It shows the proposed solution achieves close-to-optimum performance. The proposed algorithm is verified that can provide a close-to-optimum SE model selection efficiently for semantic communications.

\end{itemize}

%The remainder of the paper is organized as follows.  The system model and problem formulation is then described in Section \ref{sec:systemmodel}. Following this in Section \ref{sec:solution}, the fully polynomial-time approximation scheme using dynamic programming is proposed. In Section \ref{sec:simulation}, simulation results that demonstrate the proposed solution are given. Finally, we present the conclusions of the work in Section \ref{sec:conclusions}.

\vspace{-0.5cm}
\section{System Model and Problem Formulation}
\label{sec:systemmodel}

\begin{figure}[t]
  \centering
  \includegraphics[height=32mm, width=80mm]{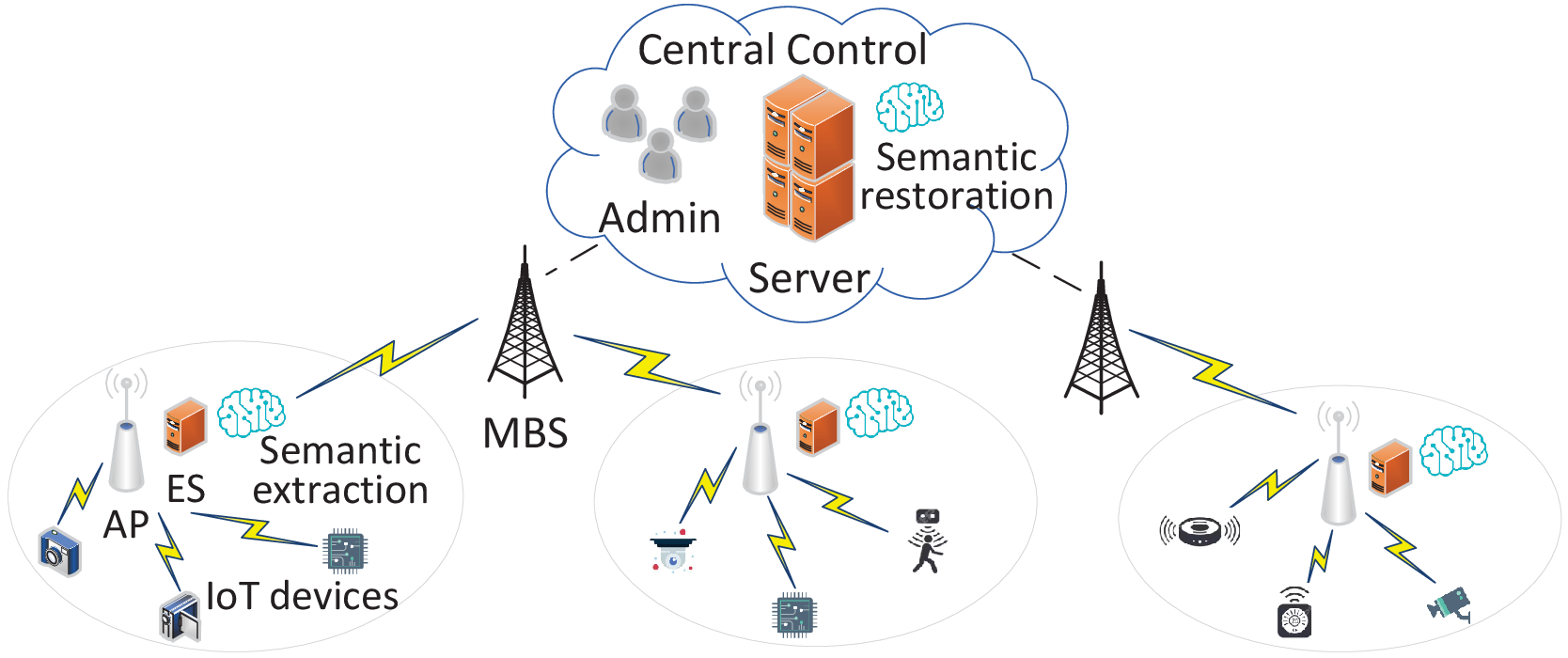}
  \caption{Edge-assisted semantic-aware network for IoT devices.}
  \label{fig:1}
  \vspace{-0.6cm}
\end{figure}

We consider an IoT network where multiple IoT devices are associated with an AP co-located with an ES. Each AP can interact with its associated macro base station (MBS), and the MBS transmits the information data to the control center through the core network for further operations, as shown in Fig. \ref{fig:1}. In order to reduce the communication loads across the whole network, semantic communications are adopted to achieve this goal. We assume that the ES can help IoT devices to complete semantic extraction and encoding. Then, the encoded semantic messages can be transmitted to the destination at the control center, which will complete the semantic restoration and decoding to give effective information.
As serving a large set of IoT devices, the computation capacity of an ES is powerful but constrained to meet diverse QoS requirements of all devices. Hence, we focus on a single AP co-located with an ES serving multiple IoT devices in the system.
Let $I$ be the total number of IoT devices in the coverage of the AP. Each device has an SE task to be executed for further semantic communications. Distinct tasks may belong to different classes, such as text-based SE, image-based SE, and goal-oriented SE.
There are $J$ SE task classes in the system. Define $\alpha_{i,j} \in \{0,1\}$ as an indicator variable representing if SE task $i$ of IoT device $i$ belongs to class $j$: if $\alpha_{i,j}=1$, SE task $i$ belongs to class $j$, otherwise $\alpha_{i,j}=0$. In general, each SE task belongs to one and only one task class, i.e., $\sum_{j=1}^{J} \alpha_{i,j} =1, \forall i$. Each IoT device uploads its SE task to the ES for execution. The ES has $K_j$ SE models to be chosen for tasks of class $j$, each requiring different amount of computing resource at the ES and achieving different SE accuracy and semantic transmission rate.
Let ${\cal I, J}, {\cal K}_j$ be the sets of IoT devices/SE tasks, task classes, and SE models for class $j$, respectively.

Let $x_{i,j,k} \in \{0,1\}$ be the decision variable representing if the SE task $i$ of class $j$ selects model $k$ to implement semantic extraction. Note that $x_{i,j,k}=0$ if $\alpha_{i,j}=0$. We assume that the SE task $i$ of class $j$ requires one and only one model in set ${\cal K}_j$ at the ES to complete the execution, i.e.,
%\vspace{-0.25cm}
\begin{equation}  \label{eq:unique}
\sum\limits_{k= 1}^{K_j} x_{i,j,{k}} = 1,\ \ \forall i,j:\alpha_{i,j}=1.
\end{equation}
Define $\varphi_{i,j,k}$ as the achieved SE accuracy for task $i$ of class $j$ using SE model $k$ and $\xi^{\min}_{i,j}$ as the minimum SE accuracy requirement of task $i$ of class $j$. Hence, the achieved SE accuracy for task $i$ of class $j$ should be no less than its minimum accuracy requirement, i.e.,
%\vspace{-0.2cm}
\begin{equation}   \label{eq:accuracy}
\sum\limits_{k= 1}^{K_j} x_{i,j,k} \varphi_{i,j,k} \ge \xi^{\min}_{i,j}.
\end{equation}
%\vspace{-0.25cm}

The total completion time of each SE task includes the wireless uploading time between IoT device and the ES and the execution time at the ES. Define $s_{i,j}$ as the input data of SE task $i$ of class $j$, which is the raw data collected by IoT device $i$ such as the pictures of the industrial environments and the videos from the smart home monitors. Let $R_i$ be the wireless uploading rate between IoT device $i$ and the AP, which can be calculated as
%\vspace{-0.3cm}
\begin{equation}   \label{eq:Ri}
R_i = w{\log _2}( 1 + \frac{{P^{\rm{T}}_i}{g _i}}{\sigma^2} ),
\end{equation}
where $w$ denotes the channel bandwidth, $g_i$ and $P^{\rm{T}}_i$ are the link gain and the transmission power from IoT device $i$ to the AP, correspondingly, and $\sigma^2$ represents the noise power at the AP receiver input.
Thus, the wireless transmission time from IoT device $i$ to the AP can be formulated as
%\vspace{-0.15cm}
\begin{equation}   \label{eq:tw}
t^{\text{w}}_{i,j}=\frac{s_{i,j}}{R_i}.
\end{equation}
Define $F$ as the maximum available computing resource of ES in number of CPU cycles/s and $f_{i,j,k}$ as the required number of CPU cycles to execute SE task $i$ of class $j$ using model $k$ at the ES. We define $t^{\max}_{i,j}$ as the maximum delay tolerance of SE task $i$ of class $j$. In order to complete the SE tasks within their delay tolerances, the following ES computation capacity constraint must be satisfied:
%\vspace{-0.2cm}
\begin{equation}  \label{eq:CPU}
\sum\limits_{i = 1}^I {\sum\limits_{j = 1}^J {\sum\limits_{k = 1}^{K_j} \frac{x_{i,j,k} f_{i,j,k}} {t^{\max}_{i,j}-t^{\text{w}}_{i,j}} } } \le F.
\end{equation}
%\vspace{-0.2cm}

The achieved semantic transmission rate after SE task $i$ of class $j$ is executed using model $k$, is denoted by $\gamma_{i,j,k}$. Unlike the bit-stream data rate, the semantic unit (sut) as the basic unit of semantic information can be used to measure the semantic information \cite{textsemantic}. Thus, semantic rate, as one of the critical semantic-based performance metrics, is defined as the transmitted semantic information per second in suts/s.
We aim to maximize the achieved total semantic rate of all SE tasks, by selecting appropriate SE models while considering SE task completion time and ES capacity constraints, and the SE accuracy requirements of IoT devices. Therefore, the SE model selection problem is formulated as an integer programming:
%\vspace{-0.15cm}
\begin{subequations} \label{Eq:maxtotal}
\begin{align}
\max_{x_{i,j,k}} &  \sum\limits_{i = 1}^I \sum\limits_{j = 1}^J \sum\limits_{k = 1}^{K_j} x_{i,j,k} \gamma_{i,j,k}    \\
\text{s.t.}~ & \eqref{eq:unique},\eqref{eq:accuracy},\eqref{eq:CPU},  \label{Eq:C1} \\
%\sum\limits_{k= 1}^{K_j} x_{i,j,{k}} = 1, ~~\forall i,j:\alpha_{i,j}=1
%&\sum\limits_{k= 1}^{K_j} x_{i,j,k} \varphi_{i,j,k} \ge \xi^{\min}_{i,j}, ~~\forall i,j \label{Eq:C2} \\
%
%&\sum\limits_{i = 1}^I {\sum\limits_{j = 1}^J {\sum\limits_{k = 1}^{K_j} \frac{x_{i,j,k} f_{i,j,k}} {t^{\max}_{i,j}-t^{\text{w}}_{i,j}} } } \le F  \label{Eq:C3} \\
%
&x_{i,j,k} \in \{0,1\}, ~~\forall i,j,k.   \label{Eq:C4}
\end{align}
\end{subequations}
%where \eqref{Eq:C1} represents that each task can choose one and only one SE model for execution; \eqref{Eq:C2} guarantees the SE accuracy meets the minimum accuracy requirement; \eqref{Eq:C3} ensures that the SE tasks are completed within the delay tolerances and the total required computation capacity does not exceed the maximum capacity of ES.
Note that problem \eqref{Eq:maxtotal} is a modification of Knapsack problem, which is a well-known NP-complete problem. Thus, it is difficult to solve it optimally in polynomial time. In Section \ref{sec:solution}, we will propose an efficient approximation algorithm to solve it with guaranteed near-optimum performance.

%%\renewcommand\arraystretch{1.3}
%%\begin{table}[htbp]
%%\begin{center}
%%\caption{Summary of Notation}
%%\label{Notation}
%%\begin{tabular}{|m{0.09\columnwidth}|m{0.53\columnwidth}|m{0.2\columnwidth}|}
%%\hline
%%Notation & Definitions & Units \\
%%\hline
%%$\mathcal{I}$  &  Set of PSs, $|\mathcal{I}|=N$ &\\
%%\hline
%%$\mathcal{K}$ &  Set of features, $|\mathcal{K}|=K$  &\\
%%\hline
%%$\mathcal{M}_k$ & Set of models for feature $k$, $|\mathcal{M}_k|=M_k$ & \\
%%\hline
%%$\beta_{i,k}$   &  Demand of PS $i$ for feature $k$ &\\
%%\hline
%%$T_{i,k}$ & Data refreshing period for feature $k$ of PS $i$ & sec\\
%%\hline
%%$\Phi_{i,k,m}$ &  Achieved accuracy for feature $k$ of DT $i$ using model $m$ &  \\
%%\hline
%%$s_{i,k,m}$ &  Amount of input data required by model $m$ to realize feature $k$ of DT $i$ & bits\\
%%\hline
%%$f_{i,k,m}$ &  Computation load needed by DT $i$ to process feature $k$ using model $m$ & CPU cycles\\
%%\hline
%%$x_{i,k,m}$ &  Decision variable indicating whether DT $i$ uses model $m$ for feature $k$  &\\
%%\hline
%%$\Psi_{i,k}$ &  Achieved accuracy for feature $k$ of DT $i$ & \\
%%\hline
%%$R_i$ & Data transmission rate of PS $i$ &  bits/s\\
%%\hline
%%$F$ & Computation capacity of the ES  & CPU cycles/sec\\
%%\hline
%%\end{tabular}
%%\end{center}
%%\end{table}

%\vspace{-0.15cm}
\section{Approximation Algorithm for the Problem}
\label{sec:solution}

Since problem \eqref{Eq:maxtotal} is NP-complete, we first transform it into a modified Knapsack problem equivalently and then propose an efficient polynomial-time approximation algorithm, which achieves a guaranteed near-optimum solution to problem \eqref{Eq:maxtotal}.
%with objective value $OBJ(S) \ge (1-\varepsilon) OBJ(S^*)$, where $S^*$ is the optimal solution to problem \eqref{Eq:maxtotal} and $\varepsilon >0$ is any given constant, while satisfying all the constraints.

%\vspace{-0.2cm}
\subsection{Problem Transformation}
\label{sec:transformation}

\begin{algorithm}[t]
\caption{Optimal solution for {\sc Typed Knapsack}}  \label{algo:DP}
\begin{algorithmic}[1]
\State{Obtain $\Omega(m,V)$ for $0\!\leq\! m\!\leq\!M$ and $0\!\leq \!V\!\leq\!V_{\max}$ by \eqref{recursion}} \label{line:1}
\State {$V^* = -\infty$}
\For {$V=0:V_{\max}$} \label{line:3}
	\If {$\Omega(M,V) \le W$}
		\State {$V^* = \max\{V^*, V \}$}	
	\EndIf
\EndFor \label{line:7}
\If {$V^* = -\infty$}
	\State\Return {Infeasible}
\Else
    \State{Obtain optimal solution $\bf x^*$ by backtracking approach}
	\State\Return {$V^*$, $\bf x^*$}	
\EndIf
\end{algorithmic}
\end{algorithm}

In order to transform the problem \eqref{Eq:maxtotal} into a modified Knapsack problem, the constraint \eqref{eq:accuracy} needs to be addressed.
%by modifying the model set $\mathcal{K}_j$.
Since $\mathcal{K}_j$ is the SE model set for task class $j$ and each task belongs to one and only one class, we define $\mathcal{K}_{i,j}$ as the SE model set of task $i$ of class $j$, in which the SE accuracy of each model is higher than or equal to the minimum requirement of task $i$ of class $j$, i.e., $\mathcal{K}_{i,j}=\{ k\in \mathcal{K}_j : \varphi_{i,j,k} \ge \xi^{\min}_{i,j} \}$.
Thus, task $i$ of class $j$ using model $k \in \mathcal{K}_{i,j}$ always satisfies its accuracy requirement \eqref{eq:accuracy}. The problem \eqref{Eq:maxtotal} is transformed equivalently into the following problem:
%\vspace{-0.2cm}
\begin{subequations} \label{Eq:maxtotal1}
\begin{align}
\max_{x_{i,j,k}} &  \sum\limits_{i = 1}^I \sum\limits_{j = 1}^J \sum\limits_{k = 1}^{K_{i,j}} x_{i,j,k} \gamma_{i,j,k}     \\
\text{s.t.}~ & \sum\limits_{k= 1}^{K_{i,j}} x_{i,j,{k}} = 1, ~~\forall i,j:\alpha_{i,j}=1  \label{Eq:CC1} \\
&\sum\limits_{i = 1}^I {\sum\limits_{j = 1}^J {\sum\limits_{k = 1}^{K_{i,j}} \frac{x_{i,j,k} f_{i,j,k}} {t^{\max}_{i,j}-t^{\text{w}}_{i,j}} } } \le F  \label{Eq:CC2} \\
&x_{i,j,k} \in \{0,1\}, ~~\forall i,j,k.   \label{Eq:CC3}
\end{align}
\end{subequations}
%\vspace{-0.5cm}

%Problem \eqref{Eq:maxtotal1} can be mapped into a modification of the classic Knapsack problem.
%The classic Knapsack problem is defined as follows: Given a set of $N$ items, a value $v_i$ and size $w_i$ for each item $i$, and a knapsack of capacity $W$, find the subset of items of total size at most $W$ and maximum total value.
%Knapsack problem can be approximately solved by a polynomial-time algorithm using Dynamic Programming (DP).
We first introduce a modification of the classic Knapsack problem related to our problem, named {\sc Typed Knapsack} \cite{DT}. Assume there are $M$ types of items, $N_m$ items belong to type $m$, and each item $n$ of type $m$ is given with weight $w_{m,n}$ and value $v_{m,n}$. The goal is to select {\em exactly} one item out of each type set into the knapsack, ensuring the total selected item weight is no more than the knapsack weight limit $W$ and the total selected item value is maximized. Compared to the classic Knapsack problem, {\sc Typed Knapsack} is distinct in its type requirements.
Problem \eqref{Eq:maxtotal1} can be mapped into {\sc Typed Knapsack}: Each task $i$ of class $j$ represents a type $(i,j)$. The SE models in $\mathcal{K}_{i,j}$ are the items of type $(i,j)$. The total type number is the total number of tasks $I$ as each task belonging to which class is given. Item $k$ of type $(i,j)$ has weight $w_{i,j,k}=\frac{ f_{i,j,k}} {t^{\max}_{i,j}-t^{\text{w}}_{i,j}}$ and value $v_{i,j,k}=\gamma_{i,j,k}$. Finally, the knapsack weight limit is the maximum computing resource $F$ of ES.
In Section \ref{sec:FPTAS}, we propose a fully polynomial time approximation scheme (FPTAS) to solve {\sc Typed Knapsack} efficiently with guaranteed near-optimum performance.

\vspace{-0.5cm}
\subsection{An FPTAS for {\sc Typed Knapsack}}
\label{sec:FPTAS}

We first propose an optimal solution for {\sc Typed Knapsack} by using designed DP based on a recursion.
%of the minimum weight of items that can achieve a total value {\em exactly} equal to a given value $V$.
The notation in the above introduced {\sc Typed Knapsack} is used for simplification.

The recursion is denoted by $\Omega(m,V)$ and given in \eqref{recursion}. $\Omega(m,V)$ denotes the minimum total weight of selected items from type 1 to $m$ achieving {\em exactly} total value $V$, by means of selecting {\em exactly} one item out of each type set.
In recursion \eqref{recursion}, the base cases are cases 1 and 2 when there is no item. In case 3, if value of each item in ${\cal N}_m$ is greater than $V$, $\Omega(m,V)=\infty$; otherwise, in case 4, $\Omega(m,V)$ selects the item in ${\cal N}_m$ yielding the minimum total weight. In the case of $\Omega(m,V)=\infty$, it indicates it is infeasible to find a solution that selects exactly one item of each type from 1 to $m$ to achieve exactly total value $V$. Let $\mathcal{N}_m$ be the set of items in type $m$.
%\vspace{-0.2cm}
\begin{equation}     \label{recursion}
\Omega(m,V)\!\!=\!\! \left\{ \!\!\!
\begin{array}{ll}
  0 & \text{if } m=0, V=0, \\
  \infty & \text{if } m=0, V>0,\\
  \infty & \text{if } m>0, \min\limits_{n\in{\cal N}_m} v_{m,n} > V, \\
  \min\limits_{n\in{\cal N}_m\!:v_{m\!,\!n} \le V} \!\!\left\{w_{m,n}+\!\!\!\!\!\! \right. \\
  \left. \Omega(m\!-\!1,V\!-\!v_{m,n})\right\}\!\!\!\! & \text{if } m>0, \min\limits_{n\in{\cal N}_m} v_{m,n} \le V,
\end{array}\right.
\end{equation}
%\vspace{-0.5cm}

Hence, a DP algorithm is proposed in Algorithm~\ref{algo:DP}, where $V_{\max}=\sum_{m=1}^M \max_{n\in{\cal N}_m} v_{m,n}$. After running line~\ref{line:1}, a table with rows corresponding to types from 0 to $M$ and the columns representing the total value $V$ from 0 to $V_{\max}$, can be obtained. The optimal total value $V^*$ is the maximum total value $V$ such that $\Omega(M,V) \le W$, found at the last row of the table where $m = M$. It is shown in lines~\ref{line:3}-\ref{line:7}. After finding the maximum total value using DP, the optimal solution of selected items is obtained by backtracking approach. Especially, in our proposed DP, we obtain a table simultaneously to record each picked item when case 4 of recursion \eqref{recursion} occurs. Thus, starting from the cell representing the optimal total value, we backtrack through both tables and return the selected item of each type contributing to the maximum total value.

However, the proposed optimal DP algorithm is pseudo-polynomial due to its computation complexity $O(MN_{\max}V_{\max})$, where $N_{\max}=\max_m N_m$ and $V_{\max}$ is pseudo-polynomial on the input size. Fortunately, the well-known value-scaling of classic Knapsack problem can be applied to design an FPTAS \cite{GareyJ79} to solve {\sc Typed Knapsack} efficiently with guaranteed near-optimum performance.
%that produces an approximate solution achieving the total value smaller than the optimal total value by a factor of at most $(1-\varepsilon)$, for any constant $\eps>0$.
Algorithm~\ref{algo:FPTAS} codifies the FPTAS algorithm, where $\theta=\eps v_{\max}/M$ denotes the scaling factor, $v_{\max}$ represents the maximum item value in the optimal solution to {\sc Typed Knapsack} and $\eps \in (0,1]$ is the precision parameter.
Since the maximum item value $v_{\max}$ in the optimal solution to original problem is unknown, we need to try all the possibilities of maximum item value, which is performed by exhaustively enumerating the values of all items for all types as $v_{\max}$.
In each iteration with given $v_{\max}$, the algorithm excludes all items with values greater than the given $v_{\max}$ (line~\ref{line:Fmax}), rounds all values down into integers in a finite range $[0,\lfloor M/\eps\rfloor]$ (line~\ref{line:F1}), and running Algorithm~\ref{algo:DP} on the rounded instance (line~\ref{line:FDP}). We then keep the solution with maximum total value among these iterations.
%Note that the optimal solution to the rounded instance is the near-optimal solution to the original instance.
%This exhaustive enumeration method increases the running time of our algorithm by a factor of $O(MN_{\max})$.

\begin{algorithm}[t]
\caption{FPTAS for {\sc Typed Knapsack} }  \label{algo:FPTAS}
\begin{algorithmic}[1]
\State {$\hat V^* = -\infty$}
\For {$m'=1:M$}
	\For {$n' = 1: N_{m'}$}
		\State {$ v_{\max} = v_{m',n'}$}
		\State {Exclude item $n$ of ${\cal N}_m$ with $v_{m,n} > v_{\max}$ for all $m,n$ } \label{line:Fmax}
        \State{$\hat v_{m,n}=\left\lfloor {v_{m,n}/\theta} \right\rfloor$ for all remaining $m,n$ } \label{line:F1}
		\State {Run Algorithm~\ref{algo:DP} using $\hat v_{m,n}$, which returns $V^*$ and $\bf x^*$} \label{line:FDP}
		\State {$\hat V^* = \max\{\hat V^*, V^* \}$ and keep solution as $\bf \hat x^*$}
	\EndFor
\EndFor
\State\Return {$\hat V^*$, $\bf \hat x^*$}
\end{algorithmic}
\end{algorithm}
%\vspace{-0.2cm}

\begin{theorem}   \label{thm:1-eps}
For any given constant $\eps \in (0,1]$, the FPTAS algorithm achieves an approximate solution (when feasible) of total value at least $(1-\eps)OBJ^*$, where $OBJ^*$ is the optimal objective value, and the computation complexity is in polynomial time $O(M^4 {N_{\max}}^2/\eps)$.
\end{theorem}
\begin{proof}
Define $S^*$ as an optimal solution to {\sc Typed Knapsack} and $v_{\max}$ as the maximum item value in optimal solution $S^*$. Let $\hat{S}$ be the optimal solution on the rounded-down instance. Define $n_m, \hat{n}_m$ as the selected items of type $m$ by $S^*, \hat{S}$, respectively. Let $OBJ(\cdot)$ be the objective value achieved by the solution using the original item values. Note that $v_{\max} \le OBJ(S^*)$. So, we have:
%\vspace{-0.3cm}
\begin{align}
OBJ(\hat{S}) &\geq  \sum_m \hat{v}_{\hat{n}_m} \theta \\
 &\geq \sum_m \hat{v}_{n_m} \theta \label{eq:dpopt} \\
 &\geq \sum_m v_{n_m} - M \theta \\
 &\geq OBJ(S^*)-\eps v_{\max} \\
 &\geq (1-\eps)OBJ(S^*),   \label{eq:lbound}
\end{align}
%\vspace{-0.05cm}
where \eqref{eq:dpopt} is because that $\hat{S}$ is an optimal solution of a maximization problem on the rounded instance.
Note that $v_{\max}$ is unknown, the exhaustive enumeration is applied to try all possible values for $v_{\max}$ and the solution achieving the maximum total value is returned. Thus, it also satisfies \eqref{eq:lbound}.

Computation complexity: Since all item values are rounded into the range $[0,\lfloor M/\eps\rfloor]$, $0\leq V_{\max} \leq M^2/\eps$. The computation complexity of DP algorithm for a given $v_{\max}$ is $O(M^3 N_{\max}/\eps)$. Adding over the computation complexity $O(MN_{\max})$ of exhaustive enumeration of $v_{\max}$, a total computation complexity $O(M^4 {N_{\max}}^2/\eps)$ is obtained, which is polynomial on the size of the inputs.
\end{proof}

%After transforming problem \eqref{Eq:maxtotal1} into {\sc Typed Knapsack}, we run the FPTAS algorithm on the resulting {\sc Typed Knapsack} instance, and that provides us with an approximate solution to the original problem, according to Theorem~\ref{thm:1-eps}.
According to Theorem~\ref{thm:1-eps}, the proposed algorithm can generate an SE model selection efficiently with guaranteed near-optimum total semantic rate in polynomial time, when the ES is requested to provide multiple models supporting the SE of multiple IoT devices with diverse QoS requirements in semantic communications.

%\vspace{-0.3cm}
\section{Simulation Results}
\label{sec:simulation}

In this section, simulation results are presented to demonstrate the superior performance of proposed solution. For comparison, the optimum solutions are obtained from exhaustive search method.
%Due to the exponential running time of optimum solution, the comparison is performed for small size systems.
In the simulation, we consider there are 6 IoT devices uniformly distributed in the coverage of an AP, which is a circular area with a radius of 150 meter. The ES is co-located at the AP and provides multiple SE models for SE tasks of associated IoT devices.
Each device generates an SE task to be executed, and each task belongs to one and only one task class. There are 4 task classes in the system and each task in each class can be executed by one of the 10 models of that class (with different semantic rate and SE accuracy).
Both the path loss and small-scale fading are considered in the link gains, given as $g_i=10^{-3}{\rho_i}^2 d_i^{-2}$, where $d_i$ is the distance between IoT device $i$ and the AP and ${\rho_i}^2$ is a random variable with exponential distribution and unit mean, since $\rho_i$ is the additional Rayleigh distributed small-scale channel fading \cite{DT}.
%Note that a 30 dB average signal power attenuation is assumed at a reference distance of 1 meter.
Table~\ref{parameters} summarizes the default parameters, where the parameter values refer to \cite{computing} and \cite{Li2023} and $U[a,b]$ represents the uniform distribution between $a$ and $b$.
%and varied during the simulation.
The simulation results are obtained by averaging over 100 independent experiments, each of which is based on one set of randomly generated IoT device locations, and task and SE model parameters.

\begin{table}[tbp]
\begin{center}
\caption{Default Parameter Settings}
\label{parameters}
\begin{tabular}{|c|c|}
\hline
Parameters     &     Values  \\
\hline
$s_{i,j}$ &  $U[2,200]$ M bits \\
\hline
$\xi^{\min}_{i,j}$ &  $U[0.65,0.8]$ \\
\hline
$t^{\max}_{i,j}$ &  $U[1200,2000]$ ms\\
\hline
$f_{i,j,k}$ &  $U[5, 500]$ M CPU cycles\\
\hline
$\gamma_{i,j,k}$ &  $U[50,200]$ M suts/s \\
\hline
$\varphi_{i,j,k}$ &  $U[0.7,1]$  \\
\hline
$F$   &  3 G Hz\\
\hline
$P_i^{\rm{T}}$ &  0.1 W\\
\hline
$w$ & 10 M Hz \\
\hline
$\sigma^2$ &  -120 dBm\\
\hline
\end{tabular}
\end{center}
\end{table}
%\vspace{-0.3cm}

\begin{figure}[t]
  \centering
  \includegraphics[width=68mm]{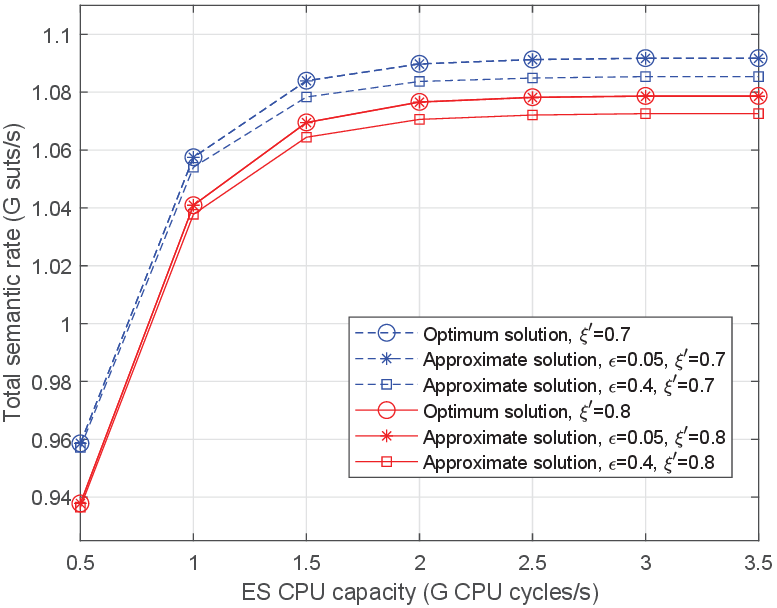}
  \caption{Total semantic rate versus ES computation capacity.}
  \label{fig:2}
  %\vspace{-0.4cm}
\end{figure}

Fig. \ref{fig:2} shows the total semantic transmission rate of all IoT devices versus the maximum ES computing resource $F$. It shows that the total semantic rate achieved by the approximate solution is close to that obtained from the optimum solution.
%We compare the achieved maximum total semantic rate using the proposed FPTAS solution and optimum solution, respectively. which demonstrates the good performance of our proposed method
Especially, the obtained approximate solutions when precision parameter $\eps=0.05$ are extremely closer to the optimum solutions than that when $\eps=0.4$. Nevertheless, the solutions obtained when $\eps=0.4$ are still close to the optimum and the running time is shorter than that when $\eps=0.05$, proven by the derived computation complexity of proposed solution in Theorem~\ref{thm:1-eps}. It further verifies the effectiveness of our proposed solution.
In addition, we can see that the achieved total semantic rate increases significantly with the ES computation capacity when $F$ is relatively small and then becomes a constant. It is because the ES computation capacity will not affect the performance anymore. It can be seen that the total semantic rate obtained from both solutions increases with the decrease of largest SE accuracy requirement $\xi'=\max_{i,j} \xi^{\min}_{i,j}$, since more models with higher semantic rate but lower SE accuracy can be chosen.

\begin{figure}[t]
  \centering
  \includegraphics[width=67.8mm]{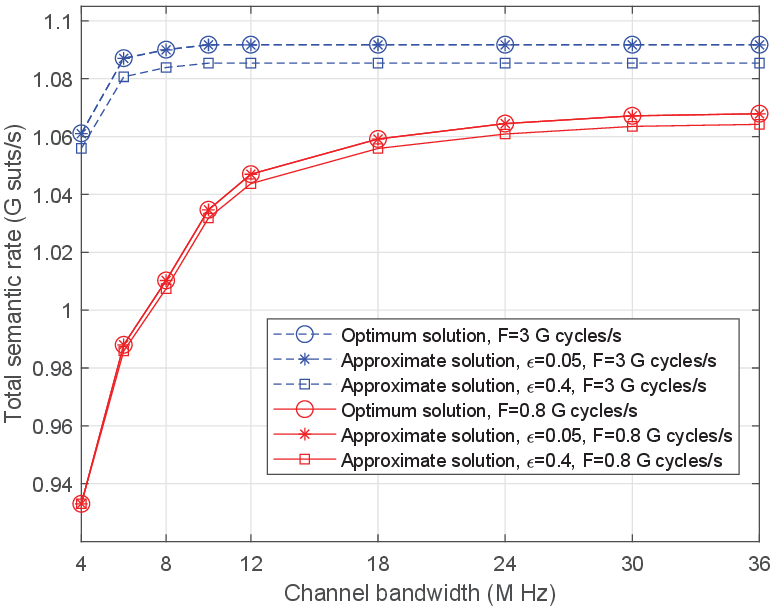}
  \caption{Total semantic rate versus wireless channel bandwidth.}
  \label{fig:3}
  %\vspace{-0.2cm}
\end{figure}

\begin{figure}[t]
  \centering
  \includegraphics[width=68.7mm]{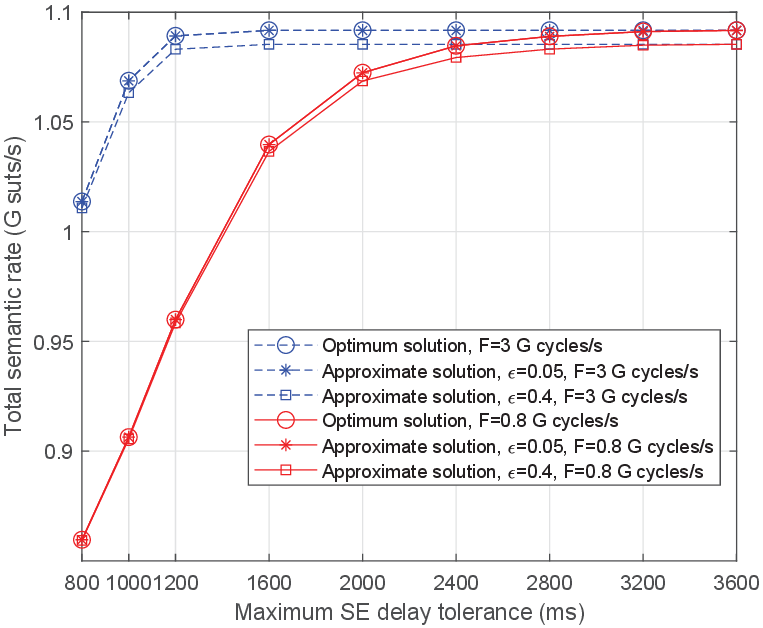}
  \caption{Total semantic rate versus maximum SE delay tolerance.}
  \label{fig:4}
  %\vspace{-0.7cm}
\end{figure}

Fig. \ref{fig:3} shows the total semantic rate versus the wireless channel bandwidth $w$. With the increase of wireless channel bandwidth, the achieved total semantic rate increases when $w$ is relatively small. The increase becomes saturated when the channel bandwidth is sufficiently large as the performance bottleneck is the ES computation capacity in this case. We can see that the total semantic rate when $F$ is 3 G cycles/s is larger than that obtained when $F$ is 0.8 G cycles/s and becomes saturated earlier. It is because when $F$ is sufficiently large, the time left for wireless raw data transmission is sufficient so that the needed wireless channel bandwidth is small.
% We can see that the total semantic rate achieved by the proposed approximate solution is close to that obtained from the optimum solution, which demonstrates the excellent performance of our proposed method.
Fig. \ref{fig:4} shows the total semantic rate versus the maximum SE delay tolerance $t^{\max}_{i,j}$ (same for all tasks). The observations are similar to fig.~\ref{fig:3}. The saturated values of total semantic rate when the maximum SE delay tolerance is large enough, are the same under different $F$. That is because when the delay tolerance is very large, the ES CPU capacity will not affect the SE model selection and the delay constraints can always be satisfied.
% However, when $t^{\max}_{i,j}$ is relatively small, the achieved total semantic rate of all IoT devices with larger ES CPU capacity is larger than that of the case with smaller ES CPU capacity. It is because larger ES computation capacity can support more SE models to be used when the tight maximum SE delay tolerance is given.
%It is also seen that the total semantic rate achieved by the proposed approximate solution is very much close to that obtained from the optimum and the gap is slightly higher when the precision parameter is bigger.
%It can be seen that the total semantic rate increases with the maximum SE delay tolerance in general when $t^{\max}_{i,j}$ is relatively small and becomes saturated as the maximum SE delay tolerance is sufficiently large.

\section{Conclusions}
\label{sec:conclusions}

In this letter, we have studied an SE model selection problem in an edge-assisted semantic-aware network, where the co-located ES and AP can support multiple SE models for various task classes associated with IoT devices. The total semantic rate of all SE tasks has been maximized by selecting appropriate SE models. Specifically, the originally formulated NP-complete integer programming problem has been transformed into a modified Knapsack problem. The proposed algorithm based on dynamic programming yields a guaranteed near-optimum solution efficiently. Simulation results have demonstrated the superior performance of proposed solution.
The proposed algorithm has been verified that can provide a close-to-optimum SE model selection efficiently for semantic communications.

%The proposed algorithm has a much lower computation complexity with close-to-optimum performance, compared to the optimal solution with exponential complexity.

\bibliographystyle{IEEEtran}
\bibliography{IEEEabrv,mybibfileSC}

\end{document}